# A Microscopically-Based, Global Landscape Perspective on Slow Dynamics in Condensed-Matter

Chengju Wang and Richard M. Stratt

Department of Chemistry
Brown University
Providence, RI 02912

**Abstract**

A complete understanding of the precipitous onset of slow dynamics in systems such as supercooled liquids requires making direct connections between dynamics and the underlying potential energy landscape. With the aid of a switch in ensembles, we show that it is possible to formulate a landscape-based mechanism for the onset of slow dynamics based on the rapid lengthening of the geodesic paths that traverse the landscape. We confirm the usefulness of this purely geometric analysis by showing that it successfully predicts the diffusion constants of a standard model supercooled liquid.

As ordinary liquids are supercooled, their dynamics changes significantly. Besides slowing down dramatically, the properties of individual degrees of freedom begin to diverge from one another, with the dynamics depending on what kinds of motions are involved and even where in the sample one looks [1-3]. This same kind of propensity for exceptionally slow and specialized dynamics is not limited to supercooled liquids; it is a common feature of disordered systems with rugged potential energy landscapes, systems ranging from proteins to glasses [4].

One way of understanding this phenomenology is to regard it as fundamentally kinetic. The *facilitated kinetics* picture derives the approach to glassy behavior by deliberately disregarding the microscopics of underlying force laws [5-9], concentrating instead on the local dynamical preferences imposed by those forces [8]. By way of contrast, "*landscape*" theories [10-14] say that the topography of the many-body potential surface is what accounts for the various kinds of behavior. It is the ruggedness or predilection towards certain geometries (e.g., the "folding funnel" portrait of protein folding [11]) that is important in these theories.

We argue here that *these approaches are not as fundamentally different as they may appear*. Much of the distinction comes from the tendency of landscape studies to extract fairly local information from potential surfaces. Indeed, traditional landscape methods focus on local minima and the saddles connecting them – a low-temperature, chemical-reaction-dynamics-like picture [15]. While these approaches are microscopic and eminently systematic, they may give insufficient conceptual weight to the dynamical difficulties of selecting the few workable routes that permit long-distance excursions through configuration space. (Equivalently, they may give insufficient weight to entropic as opposed to energetic considerations.) The facilitated kinetic models, on the other hand, do seem well suited to explaining more global phenomena. However, they are usually not directly microscopic. While they can help us explain why there are strong and fragile glasses (e.g.) [4,5], they cannot predict that molten silica will be a strong glass former, whereas $(Ca, K)(NO_3)$ will not [16].

Here we opt to concentrate on the global topological features of landscapes. To understand slow dynamics, perhaps what one should focus on are the accessible *pathways* through the landscape rather than the critical points. We suggest, in particular, that as the systems slows down, the dominant pathways are those that offer the fastest routes through an increasingly convoluted configuration space.

It is obviously difficult to find the fastest routes, because doing so would seem to require analyzing endless possible combinations of successive multi-dimensional barrier hopping processes. (Indeed the traditional focus on minima and saddles envisions individual barrier hops as the elementary ingredients in such a scheme.) However, for our purposes, we can completely remove barriers from the picture just by transforming to a different ensemble – what we call the *energy landscape ensemble*. Within this ensemble we have absolute (and not just probabilistic) definitions of the accessible points in configuration space. The allowed pathways are therefore completely characterized topologically. One can show that in this framework the long-time dynamics are completely dominated by the *shortest* of these paths, the geodesics. It is therefore possible to have a literal, microscopic, and quantitative route to predicting dynamics by analyzing those geodesics.

We can motivate our ensemble by noting that for any given total energy E, the classical trajectories followed by a many-body system will only include configurations **R** if the potential energy V(**R**) ≤ E. In fact, only systems that are ergodic will have this whole region accessible to every trajectory; nonergodic (glassy) systems will have regions that are disconnected from one another and could therefore look topologically different. It will therefore be interesting to study the connectedness, for each landscape energy $E_L$, of the ensemble of configurations **R** for which V(**R**) ≤ $E_L$. Such an approach obviously follows in the footsteps of a growing body of literature emphasizing the importance of topological features of potential energy landscapes [11-13, 17].

Our particular "landscape" ensemble differs from the conventional microcanonical ensemble in looking only at configuration space rather than phase space, and in having the energy provide an upper bound rather than an equality constraint, but it is obviously closely related. Indeed, it is possible to compute a thermodynamically equivalent configurational temperature $T_{config}$ by any one of a number of standard thermodynamic routes – from the relative configurational entropy ΔS, for example:

$$\Delta S(E_L) \equiv k_B \ln \Omega(E_L), \quad \left(T_{config}\right)^{-1} = \partial \Delta S / \partial E_L, \qquad (1)$$

with $\Omega(E_L)$ the potential energy density at the given $E_L$ and $k_B$ Boltzmann's constant. As we can see in Fig. 1, there is a 1:1 correspondence between temperature, the usual experimental control variable, and our landscape energy (outside the phase co-existence regions). For a given $E_L$, though, we no longer have to contend with activated events.

So how can we make use of this observation? Formally one can write $G(\mathbf{R} \to \mathbf{R}'|t)$, the probability of diffusing from $\mathbf{R}$ to $\mathbf{R}'$ in a time t, as a path integral weighted by an action S dependent on $V_{eff}(\mathbf{R})$, an effective potential determined by the forces acting in the problem [18, 19]

$$G(\mathbf{R} \to \mathbf{R}'|t) = \int \mathcal{D}\mathbf{R}(\tau)\, e^{-\frac{1}{2D}S[\mathbf{R}(\tau)]} \qquad (2)$$

In situations in which the diffusion constant D begins to drop precipitously, as when a liquid becomes increasingly supercooled, for example, the dominant paths are those that minimize the action. This same path-integral perspective has been exploited in a number of recent papers aimed at characterizing and finding protein-folding pathways [20, 21]. Unfortunately, the function $V_{eff}(\mathbf{R})$ is generally a complicated, temperature-dependent, quantity whose geometry may have little connection to that of any microscopic potential surface. Still, in the limiting situation suggested by the energy landscape ensemble, this action becomes quite simple. At any given $E_L$, there will be regions in configuration space $\mathbf{R}$ that are forbidden; the trajectories which move through the remaining regions will then only be governed by the portions of the potential surface for which $V(\mathbf{R}) \leq E_L$. What we suggest is that as $E_L$ decreases and $D \to 0$, *the major effect of the potential surface is embodied in the increasingly convoluted paths that the system is forced to take to avoid the energetically forbidden regions*. In particular, we hypothesize that the long-time dynamics should differ little from what one would obtain by taking the motion along the allowed regions to be completely free of external forces.

The Green's function we need then is the one for free diffusion around infinitely hard objects (Fig. 2). But the path-integral description of such a Green's function has been known for some time. Lieb pointed out that finding the exchange second virial coefficient for a hard-sphere Fermi or Bose fluid is equivalent to studying the diffusion between two points with an intervening impenetrable spherical obstacle [22]. The resulting action S depends simply on the length of the path taken between the end points, so the dominant contribution comes from the shortest such path.

$$G(\mathbf{R} \to \mathbf{R}' \,|\, t) \sim (4\pi D_0 t)^{-d/2} e^{-g^2/(4D_0 t)} , \qquad (3)$$

where g is the length of this geodesic path, $D_0$ is the force-free diffusion constant, and d is the spatial dimension. But precisely the same reasoning works for diffusion around a *series* of infinitely hard obstacles, leading to precisely same mathematical expression for the dominant contribution to (our approximation for) the relevant Green's function in the energy landscape ensemble. On the other hand, since the motion we are describing can also be regarded as free diffusion in the d = 3N dimensional configuration space, the same Green's function can be written as

$$G(\mathbf{R} \to \mathbf{R}' \,|\, t) = (4\pi D t)^{-d/2} e^{-(\mathbf{R}'-\mathbf{R})^2/(4Dt)} , \qquad (4)$$

where D is the phenomenological (experimental) diffusion constant. Comparing the two expressions implies that we should be able to predict the behavior of experimental diffusion constants from the geometry of the landscape alone, just by computing the ratio of the Euclidean and geodesic distances, $\Delta R = |\mathbf{R}' - \mathbf{R}|$ and g

$$D = \lim_{\Delta R \to \infty} D_0 \overline{(\Delta R/g)^2}, \qquad (5)$$

with the overbar representing an average over the possible end points $\mathbf{R}$ and $\mathbf{R}'$.

This argument for the dominance of landscape geodesics in slow-diffusive motion has some strong parallels with an analogous argument that has been advanced for a geometric interpretation of facilitated kinetics [8]. There the idea is that kinetic constraints effectively select a limited subspace from the whole configuration space, effectively imposing a complicated metric on the dynamics. In the limit of low temperature ($D \to 0$) this metric creates much more of a burden than any created by the specifics of the potential surface, so the dynamics is determined entirely by geometry – meaning that the optimum dynamics follows the geodesic path. But while this picture is very similar in spirit to ours, it is not as easily connected with the underlying microscopics. The space in which facilitated-kinetics takes place is assumed to be some coarse-grained version of the molecular degrees of freedom [5] so there is, at best, an indirect route from the real potential surface to the facilitation rules that are responsible for the slow dynamics. In fact, this analysis might be taken as evidence that a potential-energy-landscape perspective will not help us understand the onset of slow dynamics [6].

Our central point here is that a path-based landscape analysis can indeed provide a useful approach to obtaining a quantitative picture of slow dynamics. We illustrate that point by applying Eq. (5) to two model liquids. In both cases we take the interatomic interaction to be pair potentials of the standard truncated Lennard-Jones form,

$$u_{\alpha\beta}(r) = 4\varepsilon_{\alpha\beta}\left[\left(\sigma_{\alpha\beta}/r\right)^{12} - \left(\sigma_{\alpha\beta}/r\right)^{6}\right] - u_{\alpha\beta}^{trunc}(r) , \quad (r < r_c = 2.5\sigma_{\alpha\beta})$$

(and zero otherwise), where r is the distance between each pair of atoms of species $\alpha$ and $\beta$, $u_{\alpha\beta}^{trunc}$ is a truncation potential, and $r_c$ is the truncation distance. One example is a simple, single-component, atomic liquid, a system that undergoes a liquid-to-solid transition but does not readily supercool [23]. The other is the Kob-Andersen binary mixture, a well-studied exemplar of supercooling [24, 25].

Once we establish the connection between the landscape energy $E_L$ and configurational temperature T for each system, we need to find a representative set of path endpoints belonging to the landscape ensemble for each $E_L$. We can do so without carrying out any dynamical calculations, but here we use a (microcanonical) molecular dynamics simulation for the corresponding T to find pairs of points (**i**, **f**) in the 3N-dim configuration space separated by a prescribed Euclidean distance ΔR. Finding the true geodesic path between these pairs is a much more serious computational problem. However, finding the shortest path in our ensemble is mathematically equivalent to finding the shortest route around a set of infinitely hard obstacles sitting in an otherwise empty space – the classic path-planning problem encountered in robotics applications [26]. Using this insight it was possible to devise a number of different algorithms to locate paths that while not guaranteed to be true geodesics, at least satisfied a variety of numerical tests [25, 27].

Using these paths, we computed the diffusion constant from Eq. (5) as an average over all the (**i**, **f**) path lengths, assuming a force-free diffusion constant of the form $D_0 = \mu T$, with $\mu$ an unknown (but temperature-independent) bare mobility constant. The results are shown in Fig. 3 and compared to the exact diffusion constants computed by numerically integrating velocity autocorrelation functions derived from molecular dynamic simulations [25].

The agreement between the exact dynamics and the non-dynamical, purely geometric, predictions of the landscape theory is impressive. The overall scale of each D (the mobility constant) is set by the high temperature values, but the two-order-of-magnitude drop seen in the supercooled binary mixture near its mode-coupling temperature is a prediction that comes solely from the properties of the geodesic paths through its landscape. The particular calculations shown here are not refined enough to comment on the precise functional form of the slowing down [27]. Nor do these results provide any evidence that what we are seeing is *not* facilitated kinetics. However, the results do make it clear that one can make useful connections between the full molecular details of potential energy landscape and the advent of slow dynamics – it is just that such connections may rely more heavily on the global features of the landscape than on the local minima, maxima, and saddles.

We thank the NSF for supporting grants CHE-0518169 and CHE-0131114.

# Figure Captions

**FIG. 1.** The relation between temperature T and landscape energy $E_L$ for our atomic and binary (Kob-Andersen) liquids. Each panel presents the canonical-ensemble-average potential energy as a function of temperature (points) and the calculated configurational temperature as a function of landscape energy (lines). The Kob-Andersen liquid exhibits no thermodynamic phase transition, but the liquid-solid transition in the monatomic liquid shows up in the canonical ensemble as a flat phase co-existence region (shaded) and in the energy landscape ensemble as a jump in configurational temperature. Landscape error bars are smaller than the plotted points.

**FIG. 2.** Our landscape ensemble **(a)** and the mapping into a set of impenetrable objects **(b)** for a given landscape energy $E_L$. The forbidden parts of configuration space ($V(\mathbf{R}) > E_L$) are shaded. The geodesic route between configurations **i** and **f** is shown in **(b)** as a thick black line.

**FIG. 3** (color online). Reduced diffusion constants $D^* = D\sqrt{m/\varepsilon\sigma^2}$ for our liquids as a function of temperature T. Molecular-dynamics derived diffusion constants (points) are compared with landscape geodesic predictions (lines) for the atomic system, **(a)**, and for both the A and B particles in the Kob-Andersen binary case, **(b)** and **(c)**, (with Fig. 1 used to translate $E_L$ into T). Reliable geodesics are not found below the coexistence (shaded) region in **(a)** or the mode coupling temperature (arrow) in **(c)**. Error bars are shown for the landscape results only.

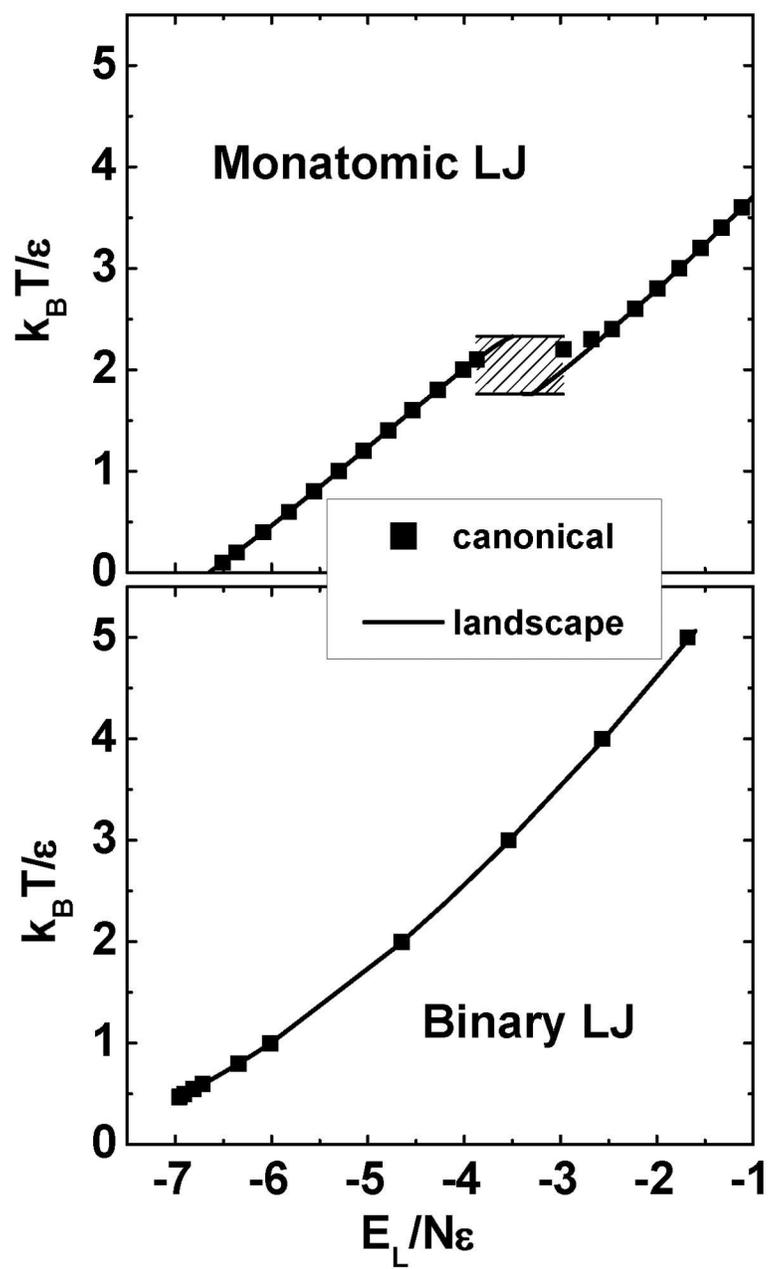

Figure 1

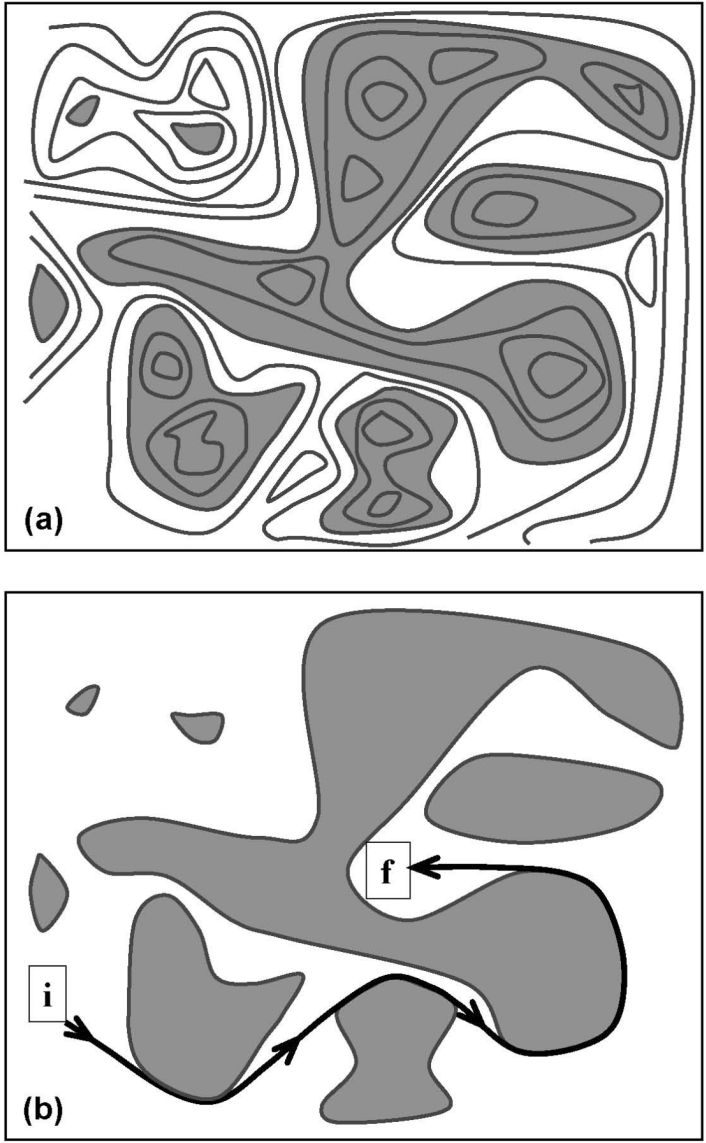

Figure 2

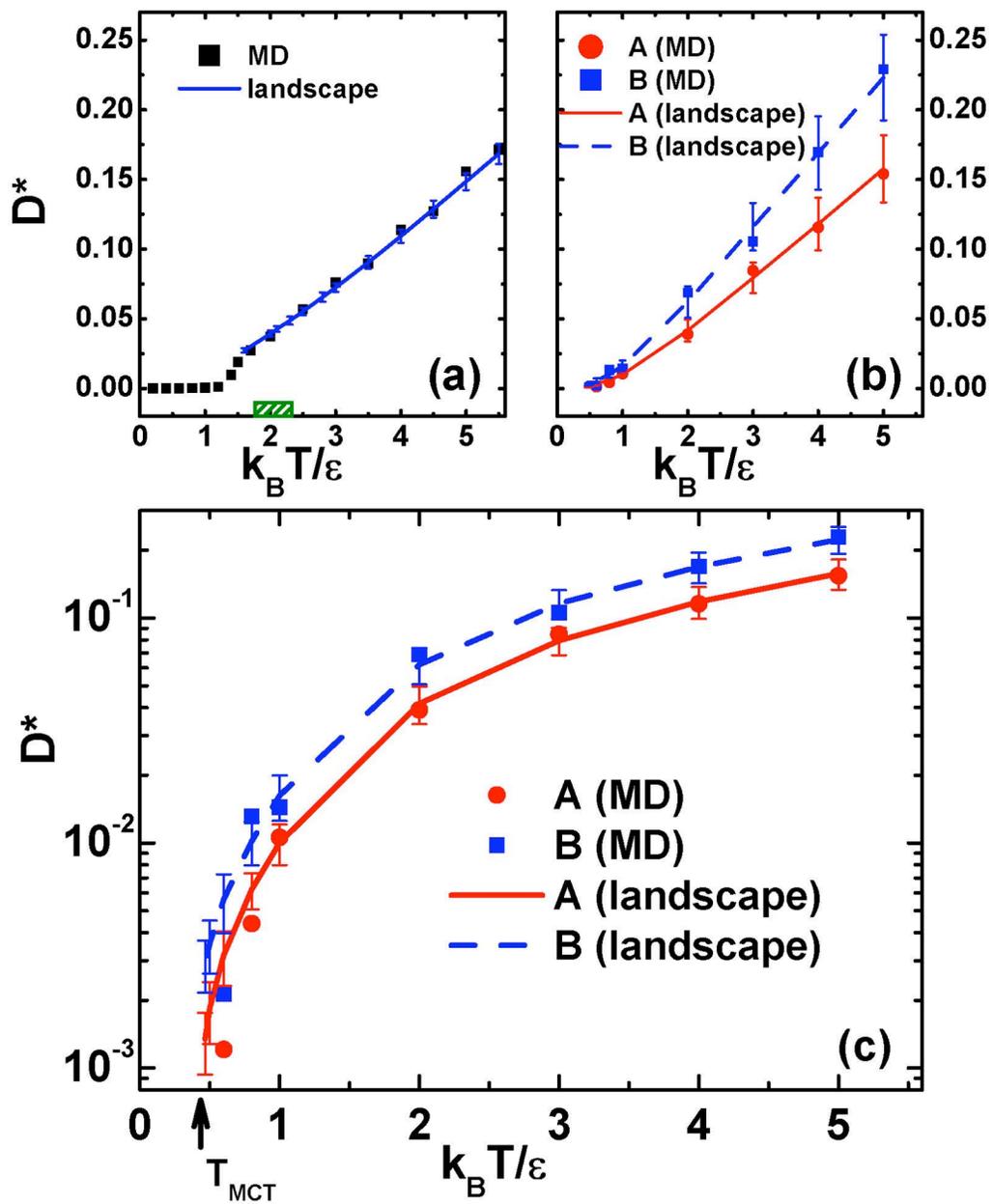

Figure 3